\documentclass[12pt]{article}
\usepackage{graphicx}
\usepackage[abs]{overpic} 

\newcommand\pubdate{\today}
\newcommand{\comment}[1]{}

\def\Title#1{\begin{center} {\Large #1 } \end{center}}
\def\Author#1{\begin{center}{ \sc #1} \end{center}}
\def\Address#1{\begin{center}{ \it #1} \end{center}}

\newcommand\pubblock{\rightline{\begin{tabular}{l} 
         \pubdate  \end{tabular}}}
\newenvironment{Abstract}{\begin{center}{\bf Abstract}\end{center} \bigskip \begin{quotation}  }{\end{quotation}}
\newenvironment{Presented}{\begin{quotation} \begin{center} 
             PRESENTED AT\end{center}\bigskip 
      \begin{center}\begin{large}}{\end{large}\end{center} \end{quotation}}

\def\beq{\begin{equation}}

\def\beqa{\begin{eqnarray}}
\def\eeqa#1{\label{#1}\end{eqnarray}}
\def\eeqan{\end{eqnarray}}

\let\bar=\overbar

\def\L{{\cal L}}

\def\Dslash{\not{\hbox{\kern-4pt $D$}}}
\def\dslash{\not{\hbox{\kern-2pt $\del$}}}

\def\msb{{\bar{\ssstyle M \kern -1pt S}}}

\def \rightdownarrow
 {\kern.3em
 \rule[.5ex]{.15mm}{2ex}
 {\mbox{$\kern-0.1em{\longrightarrow}$}}      
 }

\def\lessim{\mathrel {\vcenter {\baselineskip 0pt \kern 0pt  
\hbox{$<$} \kern 0pt \hbox{$\sim$} }}}

\def\gessim{\mathrel {\vcenter {\baselineskip 0pt \kern 0pt   
\hbox{$>$} \kern 0pt \hbox{$\sim$} }}}

\newcommand{\pipi}{\ensuremath{\pi\pi}}

\newcommand{\mpipi}{\ensuremath{m_{\pipi}}}

\newcommand{\Lumi}{\ensuremath{\mathcal{L}}}			

\newcommand{\lumifb}{\mbox{fb$^{-1}$}}				

\newcommand{\br}{\ensuremath{\mathcal{B}}}

\newcommand{\tev}{\ensuremath{\mathrm{Te\kern -0.1em V}}}
\newcommand{\gev}{\ensuremath{\mathrm{Ge\kern -0.1em V}}}	
\newcommand{\mev}{\ensuremath{\mathrm{Me\kern -0.1em V}}}	
\newcommand{\kev}{\ensuremath{\mathrm{ke\kern -0.1em V}}}	
\newcommand{\massgev}{\mbox{\gev/$c^2$}}			
\newcommand{\massmev}{\mbox{\mev/$c^2$}}			
\newcommand{\pgev}{\mbox{\gev/$c$}}				

\newcommand{\stat}{\ensuremath{\mathit{~(stat.)}}}		
\newcommand{\syst}{\ensuremath{\mathit{~(syst.)}}}		

\newcommand{\CP}{{\rm CP}}                                            

\newcommand{\pt}{\ensuremath{p_{\rm{T}}}}			

\newcommand{\ptb}{\ensuremath{\pt(B)}}				

\newcommand{\bd}{\ensuremath{B^{0}}}				
\newcommand{\bs}{\ensuremath{B^{0}_s}}				

\newcommand{\bu}{\ensuremath{B^{+}}}				

\newcommand{\bhadron}{\mbox{$b$-hadron}}			

\newcommand{\bn}{\ensuremath{B^{0}_{(s)}}}			

\newcommand{\bhh}{\ensuremath{\bn \to h^{+}h^{'-}}}
\newcommand{\fullbhh}{\ensuremath{B \to hh'}}

\newcommand{\bdkpi}{\ensuremath{\bd \to K^+ \pi^-}}

\newcommand{\bskpi}{\ensuremath{\bs \to K^- \pi^+}}

\newcommand{\bskk}{\ensuremath{\bs \to  K^+ K^-}}

\newcommand{\bspipi}{\ensuremath{\bs \to  \pi^+ \pi^-}}

\newcommand{\bdkk}{\ensuremath{\bd \to  K^+ K^-}}

\newcommand{\bsjpsiphi}{\ensuremath{\bs \to  \jpsi \phi}}

\newcommand{\bsphiphi}{\ensuremath{\bs \to  \phi \phi}}
\newcommand{\bskstarkstar}{\ensuremath{\bs \to K^{*0}\bar{K}^{*0} }}

\newcommand{\jpsi}{\ensuremath{J/\psi}}

\newcommand{\fig}[1]{fig.~\ref{fig:#1}}

\newcommand{\tab}[1]{tab.~\ref{tab:#1}}

\newcommand{\dedx}{\ensuremath{\mathit{dE/dx}}}

\newcommand{\ptot}{\ensuremath{p_{\rm{tot}}}}

\newcommand{\acp}{\ensuremath{{\cal A}_{\CP}}}
\newcommand{\acpbdkpi}{\ensuremath{\acp(\bdkpi)}}

\newcommand{\etal}{et al.}

\def\babar{\mbox{\slshape B\kern-0.1em{\smaller A}\kern-0.1em B\kern-0.1em{\smaller A\kern-0.2em R}}}
  
\newcommand{\belle}{Belle}
\newcommand{\BaBar}{BaBar}

\newcommand{\Bdpipi}{\ensuremath{\bd \rightarrow \pi^+ \pi^-}}

\newcommand{\BdKpi}{\ensuremath{\bd \rightarrow K^{+} \pi^-}}

\newcommand{\BsKpi}{\ensuremath{\bs \rightarrow K^- \pi^+}}

\newcommand{\BsKK}{\ensuremath{\bs \rightarrow  K^+ K^-}}

\newcommand{\Bspipi}{\ensuremath{\bs \rightarrow  \pi^+ \pi^-}}

\newcommand{\BdKK}{\ensuremath{\bd \rightarrow  K^+ K^-}}

\newcommand{\Lbppi}{\ensuremath{\Lambda_{b}^{0} \rightarrow p\pi^{-}}}

\newcommand{\LbpK}{\ensuremath{\Lambda_{b}^{0} \rightarrow pK^{-}}}

\newcommand{\BuKpi}{\ensuremath{\bu \rightarrow K^{+} \pi^0}}

\newcommand{\BR}{\ensuremath{\mathcal B}}

\newcommand{\BspipisuBdKpidef}{\ensuremath{\frac{\mathit{f_s}}{\mathit{f_d}}\times\frac{\BR(\Bspipi)}{\BR(\BdKpi)}}}
\newcommand{\BdKKsuBdKpidef}{\ensuremath{\frac{\BR(\BdKK)}{\BR(\BdKpi)}}}

\def\beq{\begin{equation}}
\def\eeq{\end{equation}}
\def\bea{\begin{eqnarray}}
\def\eea{\end{eqnarray}}

\def\sss{\scriptscriptstyle}

\def\barp{{\raise.35ex\hbox
{${\sss (}$}}---{\raise.35ex\hbox{${\sss )}$}}}
\def\bdbarp{\hbox{$B_d$\kern-1.4em\raise1.4ex\hbox{\barp}}}
\def\bsbarp{\hbox{$B_s$\kern-1.4em\raise1.4ex\hbox{\barp}}}

\def\roughly#1{\mathrel{\raise.3ex\hbox
{$#1$\kern-.75em\lower1ex\hbox{$\sim$}}}}

\newcommand{\Lxy}{\ensuremath{L_{\rm{T}}}}			

\newcommand{\acpbskpi}{\ensuremath{\acp(\bskpi)}}

\newcommand{\Bd}{\ensuremath{B^{0}}}

\newcommand{\Bu}{\ensuremath{B^{+}}}
\newcommand{\Bs}{\ensuremath{B_{s}^{0}}}

\newcommand{\Lb}{\ensuremath{\Lambda_{b}^{0}}}

\newcommand{\bear}{\begin{array}}
\newcommand{\ear}{\end{array}}
\newcommand{\bet}{\begin{tabular}}
\newcommand{\eet}{\end{tabular}}
\newcommand{\beqn}{\begin{eqnarray}}
\newcommand{\eeqn}{\end{eqnarray}}

\newcommand{\Dkpi}{\ensuremath{D^{0} \rightarrow K^- \pi^+}}

\newcommand{\bskzerostarkzerostar}{\ensuremath{\bs \to  K^{0*} \bar{K}^{0*}}}

\newcommand{\phiphi}{\ensuremath{B_s^0 \to \phi \phi}}
\newcommand{\phikk}{\ensuremath{\phi \to K^+K^-}}
\newcommand{\jpsiphi}{\ensuremath{B_s^0 \to J\!/\!\psi \phi}}
\newcommand{\jpsifnot}{\ensuremath{B_s^0 \to J\!/\!\psi f_0(980)}}

\newcommand{\phikst}{\ensuremath{B^0 \to \phi K^{\star}(892)^0}}

\newcommand{\au}{\ensuremath{\mathcal{A}_u}}
\newcommand{\av}{\ensuremath{\mathcal{A}_v}}

\newcommand{\cdf}{CDF Collaboration}

\newcommand{\lhcb}{LHCb Collaboration}

\begin{document}
\begin{titlepage}
\pubblock

\vfill


\Title{Measurements of B meson decay rates and \CP\ violating asymmetries}
\vfill
\Author{Fabrizio RUFFINI}  
\Address{INFN of Pisa and University of Siena\\
Polo Fibonacci, Largo B. Pontecorvo, 56127 Pisa - Italy\\
E-mail: ruffini@pi.infn.it
}
\vfill


\begin{Abstract}
Hadrons containing $b-$quarks represent a great opportunity to investigate the flavor sector of the Standard Model (SM) and to look for New Physics effects (NP). In this report we review the most up-to-date results on $b$-hadron decay rates and \CP\ violating asymmetries: among other results, we report Branching Ratio (\BR) and CP Violation (CPV) of  \Bd, \Bs\ and \Lb\ decay modes into pairs of charmless charged hadrons (pions, kaons and protons), the first search for CPV in \phiphi\ decays and the first observation of \bskzerostarkzerostar\ decay. Two new results from the CDF collaboration are reported: the first evidence of \bspipi\ decay and the world's best measurement of \BR (\bdkk).

\comment{
In this report we review the most up-to-date results 
on B meson decay rates and \CP\ violating asymmetries: among other results, we report BR and CPV of  \Bd, \Bs\ and \Lb\ decay modes into pairs of charmless charged hadrons 
(pions, kaons and protons), the first search for CPV in \phiphi\ decays and the first observation of \bskzerostarkzerostar\ decay with a 7$\sigma$ significance.  The preliminary result is \BR(\bskzerostarkzerostar) =  $1.95 \pm 0.47\stat \pm 0.51 \syst \pm 0.29 (f_{d}/f_{s}) \times 10^{-5}$.

Two new results of the CDF collaboration are reported: the first evidence of \bspipi\ decay and the world's best measurement of \bdkk\ decay.
Using a data sample corresponding to \mbox{6\lumifb} of integrated luminosity, 
CDF reported the first evidence of the \Bspipi\ decay, with a significance of 
$3.7\sigma$, and measured $\BR(\Bspipi)= (0.57 \pm 0.15\stat \pm 0.10\syst)\times 10^{-6}$. 
No evidence is found for the decay \BdKK\, and CDF set a 90\% confidence level interval  $[0.05,0.46] \times 10^{-6}$,
corresponding to the central value  $\BR(\BdKK)= (0.23 \pm 0.10\stat \pm 0.10\syst)\times 10^{-6}$.
}

\end{Abstract}

\vfill

\begin{Presented}
The Ninth International Conference on\\
Flavor Physics and CP Violation\\
(FPCP 2011)\\
Maale Hachamisha, Israel,  May 23--27, 2011
\end{Presented}
\vfill

\end{titlepage}
\def\thefootnote{\fnsymbol{footnote}}
\setcounter{footnote}{0}
%


\section{Introduction}
The interpretation of the CP violation mechanism is one of the most controversial aspects of the Standard Model.
Many extensions of Standard Model predict that there are new sources of CP violation,
beyond the single Kobayashi-Maskawa phase in the quark-mixing matrix (CKM). Considerations 
related to the observed baryon asymmetry of the Universe imply that such new sources should exist.

The non-leptonic decays of $b$ hadrons into pairs of charmless charged hadrons 
are effective probes of the CKM matrix and sensitive to potential new physics effects.
The large production cross section of $b$ hadrons of all kinds at the TeVatron and LHC
allows extending such measurements to \Bs\ and \Lb\ decays,
which are important to supplement our understanding of \Bd\  and \Bu\ meson decays provided by the $B$-factories.
The branching fraction of \BsKpi\ decay mode provides information on the CKM angle 
$\gamma$~\cite{Gronau:2000md} and the measurement of direct
\CP\ asymmetry could be a powerful model-independent test 
of the source of \CP\ asymmetry in the $B$ system \cite{Lipkin:2005pb}. 
The \Bspipi\ and \BdKK\  decay modes proceed through annihilation and exchange topologies, which
are currently poorly known and a source of significant uncertainty in
many theoretical calculations~\cite{B-N,Bspipi}. 
A measurement of both decay modes would allow a determination of
the strength of these amplitudes~\cite{Burasetal}.


The measurement of the \BsKK\ lifetime may be used to put constraints on contributions from NP to the $\bs$ mixing phase and the width difference between the light and the heavy states in the \bs\ system $\Delta \Gamma_{s}$. In addition, \BsKK\ decay being dominated by loop diagrams, new particles entering the loop processes can significantly influence \BsKK\ decay.

Present CDF statistics of the \phiphi\ data sample allows investigations of the Triple Product asymmetries, a class of CP-violation observables which can reveal the presence of NP \cite{alakabha}. 

\bskzerostarkzerostar\ is a decay into two light vector mesons that proceeds solely through loop penguin $b\to s$ diagrams within the SM. The interest in \bskzerostarkzerostar\ is related to the extraction of $\beta_{s}$ and $\gamma$ \cite{alakabha_2}.

Throughout this paper, C-conjugate modes are implied and branching fractions indicate \CP-averages unless otherwise stated. In addition, the first uncertainty associated with any number is statistical, while the second one is systematic.


\section{CDF\,II}
The Collider Detector at Fermilab (CDF\,II) experiment is a multipurpose magnetic spectrometer surrounded by
calorimeters and muon detectors~\cite{CDF}. 
A silicon micro-strip detector (SVXII) and a cylindrical drift chamber
(COT) situated in a 1.4 T solenoidal magnetic field
reconstruct charged particles in the pseudo-rapidity range
$|\eta| < 1.0$.
The transverse momentum resolution is $\sigma_{p_{T}}/p_{T} \simeq
0.15\%\, p_{T}$/(GeV/$c$) and the observed mass-widths are about 14 \massmev\
for $J/\psi\to\mu^+\mu^-$ decays, and about 9 \massmev\ for \Dkpi\ decays.
 The specific energy loss by ionization (\dedx) of charged particles in the COT
is measured from the amount of charge collected by each wire.
An average separation power of 1.4 Gaussian-equivalent standard deviation ($\sigma$) is obtained in separating 
pions and kaons with momentum larger than 2~\pgev. Consequently the separation between $KK$ and $\pi\pi$,  or $K^{+}\pi^{-}$ and $K^{-}\pi^{+}$ corresponds to about 2.0 $\sigma$.

\section{LHCb}
The LHCb detector \cite{LHCb} is a forward spectrometer covering the pseudo-rapidity range $1.8 < \eta < 4.9$, 
designed to perform flavour physics measurement at the LHC and composed of several specialized sub-systems. The tracking system consists of a vertex detector, which allows an accurate reconstruction of the primary vertex, and a set of tracking stations in front of and behind a dipole magnet that provides a field of 4 Tm.
Overall, the tracking system provides an impact parameter resolution 
of $\sim$ 16 $\mu$m + 30 $\mu$m$/p_{T}$(GeV/c), and a momentum resolution that ranges from $\sigma/p \sim$ 0.5\% at 3 GeV/c and to $\sim$ 0.8\% at 100 GeV/c.
Two RICH (Ring-Imaging Cherenkov) detectors are used together and provide a typical kaon efficiency of 
$\sim$95\% for a pion fake rate of a few percent, integrated over the momentum range from 3-100 GeV/c. 
Downstream of the second RICH are a Preshower/Scintillating Pad Detector, an electromagnetic calorimeter, a hadronic calorimeter and 5 muon chambers.


\section{First evidence of \bspipi\ decay at CDF}
The extraction of CKM parameters from measurements in $b$-hadron decays is often affected by large uncertainties, coming from non-perturbative QCD effects. One way to simplify the problem is to use flavor symmetries under which the unknown effects partially cancel. Therefore the simultaneous study of \Bd\ $\to hh$ and \Bs\ $\to hh$ decays (where $h$ can be a pion or a kaon) is particularly interesting since these modes are related by subgroups of the SU(3) symmetry.
Significant contributions from higher-order (``penguin'') transitions provide sensitivity to the possible presence of NP in internal loops, if the observed decay rates are inconsistent with expectations. Of the possible \fullbhh\ decay modes 
only the $\bspipi$ and $\bdkk$ observations are still missing. A measurement of the \br\ of the \bspipi\ mode, along with the \bdkk\ mode, would allow a determination of the strength of penguin-annihilation amplitudes, which is currently poorly known and source of significant uncertainty in many calculations.

CDF analyzed an integrated luminosity  
$\int\Lumi dt\simeq 6$~\lumifb\ sample of pairs of oppositely-charged particles
with $p_{T} > 2$~\pgev\ and   $p_{T}(1) + p_{T}(2) > 5.5$~\pgev,
used to form $B$ candidates.
The trigger required also a transverse opening angle $20^\circ < \Delta\phi < 135^\circ$ between the two tracks,
  to reject background from particle 
pairs within the same jet and from back-to-back jets.
In addition, both charged particles were required to originate from
a displaced vertex with a large impact parameter (100 $\mu$m $< d_0(1,2) < 1$~mm), 
while the \bhadron\ candidate was required to be produced in
the primary $\bar{p}p$ interaction ($d_0< 140$~$\mu$m) and to have travelled a transverse distance
$\L_{T}>200$~$\mu$m. A sample of about 3 million 
 \fullbhh\ decay modes  
(where  $ B = \Bd,\Bs ~{\rm or}~ \Lb$  and $h= K~{\rm or}~ \pi$) 
was reconstructed after the off-line confirmation of trigger requirements. 
In the offline analysis, an unbiased optimization procedure determined a
tightened selection on track-pairs fit to a common decay vertex.
%
The offline selection is based on a more accurate 
determination of the same quantities used in the trigger, with the 
addition of two further observables: 
the isolation ($I_{B}$) of the $B$ candidate~\cite{Isolation},
and the quality of the three-dimensional fit ($\chi^{2}$ with 1 d.o.f.) of the 
decay vertex of the $B$ candidate. 
Requiring a large value of $I_{B}$ reduces the background from 
light-quark jets, and a low $\chi^{2}$ reduces the 
background from decays of different long-lived particles within the event.
The final selection, inherited from Ref.~\cite{Aaltonen:2008hg}, 
was originally devised for the \BsKpi\ search, but has proven to be optimal also for
detection of  the \Bspipi\ and includes the following criteria: 
 $I_B>0.525$, $\chi^{2}<5$,
$d > 120~\mu m$, $d_B < 60~\mu m$, and $\Lxy > 350~\mu m$.
No more than one $B$ candidate per event is found after this 
selection, and a mass ($m_{\pi\pi}$) is assigned to 
each, using a charged pion mass assignment for both decay products. 

\begin{figure}[htb]
\centering
\begin{overpic}[scale=0.35]{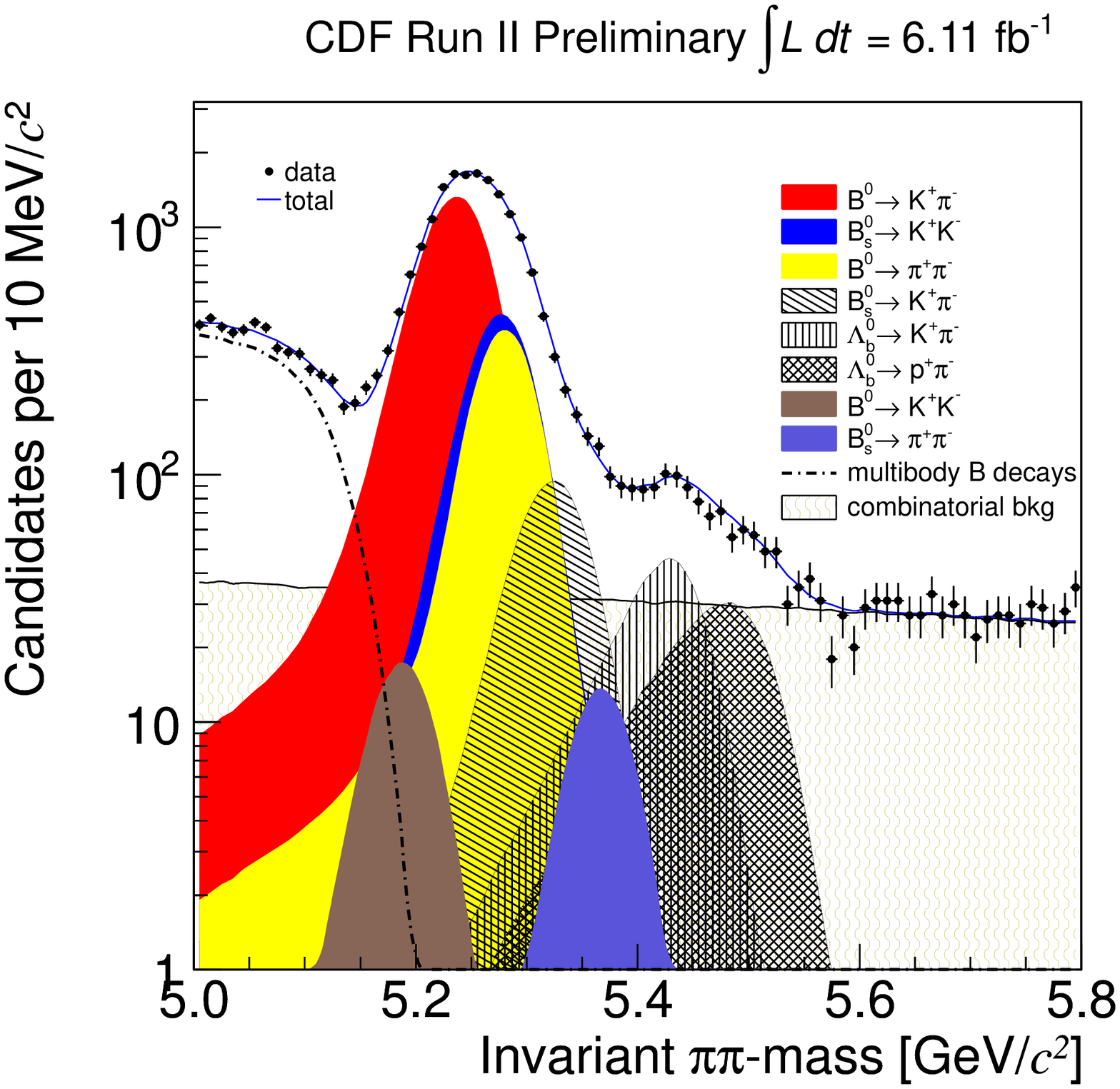}
\put(174,160){(a)}
\end{overpic}   
\begin{overpic}[scale=0.35]{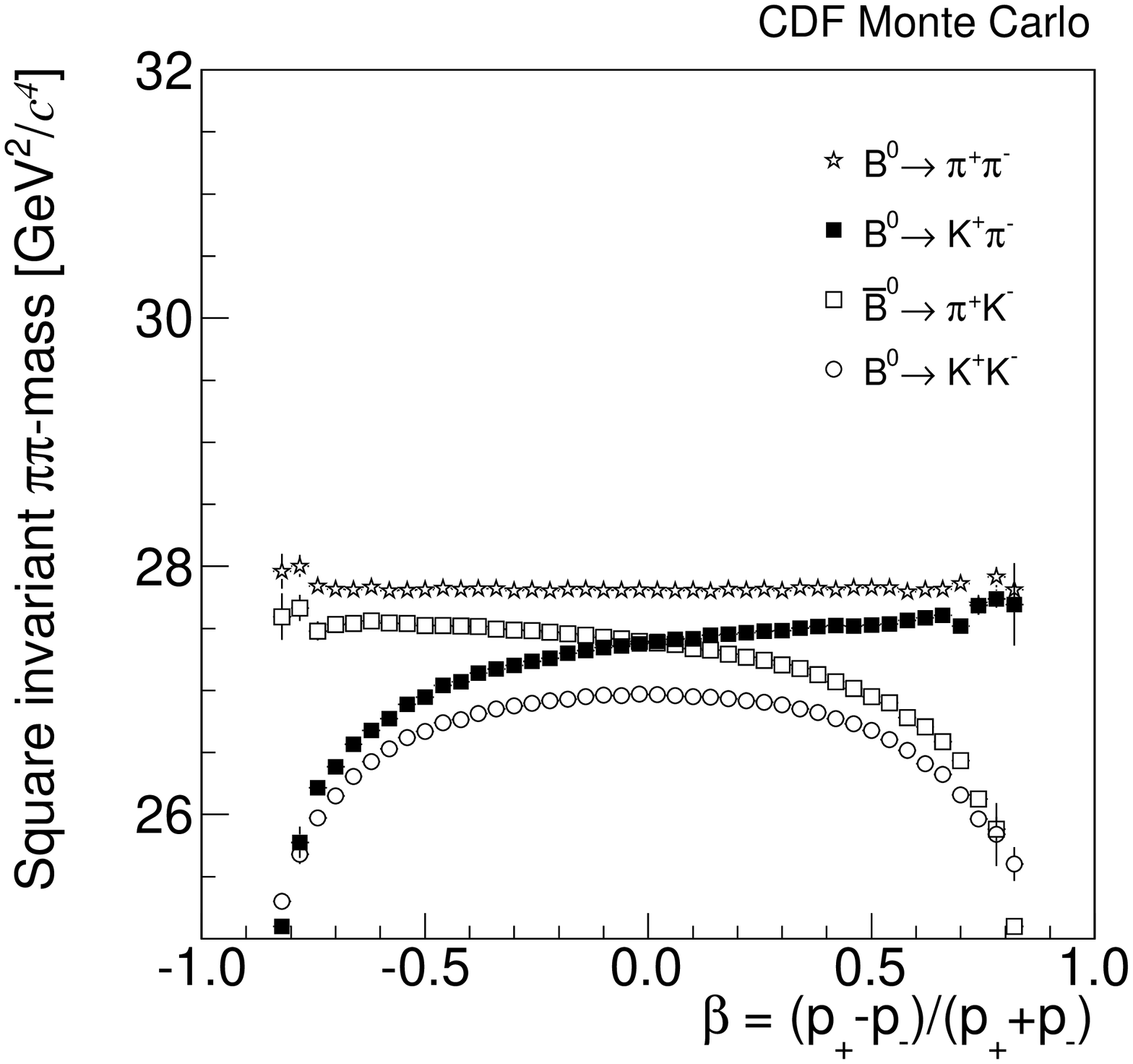}
\put(174,160){(b)}
\end{overpic}   
\caption{Invariant mass distribution of \fullbhh\ candidates passing cuts selection,  using a pion mass assumption for both decay products. Cumulative projections of the Likelihood fit for each mode are overlaid (a). Profile plots of the square invariant $\pi\pi$-mass as a function of charged momentum asymmetry $\beta$ for all \Bd\ simulated signal modes (b).} 
\label{fig:projections}
\end{figure}

\comment{
In the offline analysis the discriminating power of the $B$ isolation and of the information provided by the 3D reconstruction 
capability of the CDF tracking were also used,
allowing an improvement in the signal purity.
 Isolation is defined as $I(B)= \ptb/[\ptb + \sum_{i} \pt(i)]$, in which the sum runs over every other track 
(not from the $B$ hadron) within
a cone of unit radius in the $\eta-\phi$ space around the $B$ hadron flight direction. 
By requiring $I(B)> 0.5$,  the background was reduced by a factor
4 while keeping almost 80\% of signal. The 3D silicon tracking allowed  multiple vertices  to be resolved
along the beam direction and the rejection of fake tracks, reducing the background
by a factor of 2, with only a small efficiency loss on signal.
}
The resulting $\pi\pi$-mass distributions (see Figure~\ref{fig:projections}, (a)) show a clean signal of \fullbhh\ decays.
Backgrounds include mis-reconstructed multibody $b$-hadron decays (physics background, causing the enhancement at \mpipi $< 5.16$ \massgev) and random pairs of charged particles (combinatorial background).
In spite of a good mass resolution ($\approx 22\,\massmev$), the various \fullbhh\ modes overlap into an unresolved
mass peak near the nominal \Bd\ mass, with a width of about $\approx 35\,\massmev$. 

The resolution in invariant mass and in particle identification, provided by specific ionization energy loss  (\dedx) in the drift chamber, is not sufficient for separating the individual \fullbhh\ decay modes on an event-by-event basis,
therefore a Maximum Likelihood fit, incorporating kinematics and PID information, was performed.  The kinematic information is summarized by three loosely correlated
observables: (a) the square of the invariant $\pi\pi$ mass
$m^{2}_{\pi\pi} $; (b) the signed momentum imbalance
$\beta = (p_{+} - p_-)/(p_+ + p_{-})$, where $p_+$ ($p_-$) is the
the momentum of the positive (negative) particle; (c) the scalar sum of particle momenta 
$\ptot=p_+ + p_-$.
\comment{
The above variables allow evaluation of the square of the invariant mass $m^{2}_{12}$ of a candidate for  any mass assignment of the decay products ($m_{1}$,$m_{2}$), using the equation
\begin{eqnarray}\label{eq:Mpipi2}
 m^{2}_{12}  =  m^{2}_{\pi\pi}  -  2 m_{\pi}^2 + m_{+}^2+m_{-}^2 +
                 \nonumber    \\
          -  2 \sqrt{p_{+}^2+m_{\pi}^2} \sqrt{p_{-}^2+m_{\pi}^2}  
                    +  2\sqrt{p_{+}^2+m_{+}^2} \sqrt{p_{-}^2+m_{-}^2},
 \end{eqnarray}
}
The likelihood exploits the kinematic differences among modes (see Figure~\ref{fig:projections}, (b)) by using the correlation between the signed momenta of the tracks and the invariant masses $m^{2}_{+-}$ of a candidate for any mass assignment of the decay products ($m_{+}$,$m_{-}$), using the equation
\begin{eqnarray}\label{eq:Mpipi2}
 m^{2}_{+-}  =  m^{2}_{\pi\pi}  -  2 m_{\pi}^2 + m_{+}^2+m_{-}^2 +
                 \nonumber    \\
          -  2 \sqrt{p_{+}^2+m_{\pi}^2} \sqrt{p_{-}^2+m_{\pi}^2}  
                    +  2\sqrt{p_{+}^2+m_{+}^2} \sqrt{p_{-}^2+m_{-}^2}.
 \end{eqnarray} 
This procedure is useful to obtain statistical separation power between $\pi\pi$ and $KK$ (or $K\pi$) final states, and therefore is a key tool for the measurement of the observables of interest.
Kinematic fit templates are extracted from simulation for signal and physics background, while they are extracted from an independent sample
for combinatorial background.
Mass line-shapes are accurately described according for the non Gaussian resolution tails and for the effects  of the final state radiation of the soft photons. The \dedx\ is calibrated over the tracking volume and time using about 3.2 millions of $D^{*+} \to D^{0} [\to K^{-}\pi^{+}]\pi^{+}$ decays, where the sign of the soft pion tags the $D^{0}$ flavor. A 1.4$\sigma$ separation is obtained between kaons and pions with $p > 2$ \pgev, becoming 2.0$\sigma$ for the couples $KK/\pi\pi$ and $K^{+}\pi^{-}/K^{-}\pi^{+}$. 
 \dedx\ templates (signal and background) are extracted from the $D^{0}$ samples used in calibration.


The dominant contributions to the systematic uncertainty are the uncertainty on the \dedx\ calibration and parameterization and 
the uncertainty on the combinatorial background model. An additional systematic uncertainty, of the order of 10\% has been assessed because of a fit bias, found in the estimate of the relative fraction of the \BdKK\ decay mode. 
Other contributions come from trigger efficiencies, physics background shape and kinematics, $b$--hadron masses and lifetimes.

The signal yields are calculated from the signal fractions returned by the likelihood fit.
For the first time significant signal is seen for \Bspipi, with a significance of $3.7\sigma$, while the significance for the \BdKK\ decay mode is $2.0$ $\sigma$. 
The significances were estimated combining the statistical significance returned by the fit and a systematic uncertainty evaluated using the Likelihood Ratio distribution (distributed with good approximation as a $\chi^{2}$ distribution) on pseudo-experiments.

\begin{table}[h]
{
{
\scriptsize
\begin{tabular}{|l|l|c|l|}
\hline
\hline
Mode          & Quantity & Relative \BR & Absolute \BR (10$^{-6}$)  \\
\hline
\BdKK          & \BdKKsuBdKpidef\    	&  0.012 $\pm$ 0.005 $\pm$ 0.005   & 0.23 $\pm$ 0.10 $\pm$ 0.10 ($[0.05, 0.46]$ @~90\%~CL)  \\
\Bspipi        & \BspipisuBdKpidef\  	&  0.008 $\pm$ 0.002 $\pm$ 0.001   & 0.57 $\pm$ 0.15 $\pm$ 0.10  \\
\hline
\hline
\end{tabular}
}
}
\caption{\label{tab:summary_Bspipi} Branching fractions results \cite{cdfnote:10498}. Absolute branching fractions are normalized to the the world--average values
${\mathcal B}(\mbox{\BdKpi}) = (19.4\pm 0.6) \times 10^{-6}$ and $f_{s}/f_{d}= 0.282 \pm 0.038$~\cite{pdg_2010}.}
\end{table}

The relative branching fractions are listed in Table~\ref{tab:summary_Bspipi}, where $f_{d}$ and $f_{s}$ indicate the production fractions respectively of \Bd\ and \Bs\ from fragmentation of a $b$ quark in $\bar{p}p$ collisions.
Absolute results are also listed in Table~\ref{tab:summary_Bspipi}; they are obtained 
by normalizing the data to the world--average of \BR(\BdKpi)~\cite{pdg_2010}. 
An 90\% of confidence level interval is also quoted for the \BdKK\ mode. 

The present measurement of $\BR(\BdKK)$ is the world's best measurement and supersedes the previous limit~\cite{Abulencia:2006psa}. The central value is in agreement with other existing measurements~\cite{pdg_2010}, 
while it is higher than the predictions~\cite{Beneke:2003zv}\cite{Cheng:2009cn}.

The branching fraction of the \Bspipi\ mode is consistent with the previous upper limit ($< 1.2 \times 10^{-6}$ at 90\%~C.L.), based on a subsample of the current data \cite{Aaltonen:2008hg}, and is better than other existing measurements \cite{Peng:2010}; it is in agreement with the theoretical expectations within the pQCD approach \cite{Ali:2007ff}, \cite{Li:2004ep} while is higher than most other predictions ~\cite{Beneke:2003zv} \cite{Sun:2002rn} \cite{Chiang:2008vc} \cite{Cheng:2009mu}.

More details on the analysis can be found in \cite{cdfnote:10498}.

\clearpage
\section{Other \BR\ and CPV measurements of \fullbhh\ decays}
\bhh\ decays make possible a rich flavor-physics program at the B-factories and at the hadron machines.
The branching fraction of \BsKpi\ decay mode provides information on the CKM angle 
$\gamma$~\cite{Gronau:2000md} and the measurement of direct
\CP\ asymmetry could be a powerful model-independent test 
of the source of \CP\ asymmetry in the $B$ system \cite{Lipkin:2005pb}. 
As recently suggested from Lipkin \cite{Lipkin:2011hh}, probably solving this puzzle, a discrepancy between the direct CP asymmetries in the \BdKpi\ and \BuKpi\ decays is expected within the SM, and it is experimentally observed ~\cite{pdg_2010}. 

Nevertheless, high accuracy measurements of \acpbdkpi\ and \acpbskpi\ may provide useful information to our comprehension of this discrepancy.
CP violating asymmetries in \Lbppi\ and \LbpK\ decay modes may reach significant size $ \mathcal{O}(10\%)$ in the Standard Model \cite{Mohanta1}. 
Measurements of asymmetries and branching fractions of these modes may favor or disfavor some specific extensions or would rule out (or allow) some extensions of 
the SM \cite{Mohanta2}.  

We report a comparison of the most up-to-date measurements performed by Belle \cite{Peng:2010}, \cite{Lin:2007}, \cite{Lin:982007}, BaBar \cite{aubert:2007}, CDF \cite{Aaltonen:2008hg}, \cite{Aaltonen:2011qt}, and LHCb \cite{acp_LHCb_vagnoni} of \BR\ and CPV asymmetries of \fullbhh\ decay modes (see \tab{BR_Bd}, \tab{BR_Bs}, \tab{ACP_Bhh}).
CDF measurements on \Bd\ and \Bs\ sector are in agreement and competitive with the results from B-factories. In particular,  \acpbskpi\ the measurements for the \Lb\ sector represent the first measurement of these observables. 

\begin{table*}[!h]
{ 
\begin{tabular}{|l|rrr|}
\hline
\hline
Mode 	& CDF 			& BaBar 						& Belle \\
		&1fb$^{-1}$		&467 M ($B\bar{B}$ pairs)		& 535 M ($B\bar{B}$ pairs) \\
\hline
\BdKpi	&						&  19.1 $\pm$0.6 $\pm$ 0.6	& 19.9 $\pm$ 0.4$\pm$ 0.8  \\
\Bdpipi	& 5.02 $\pm$ 0.33 $\pm$ 0.35 	& 5.5 $\pm$ 0.4 $\pm$ 0.3		& 5.1 $\pm$ 0.2 $\pm$ 0.2\\
\BdKK   	& 0.39 $\pm$ 0.16 $\pm$ 0.12    	& 0.04 $\pm$ 0.15 $\pm$ 0.08	& 0.09 $^{+0.18}_{-0.13}$ $\pm$0.01 \\
\hline
\hline
\end{tabular}
}
\caption{\br\ (10$^{-6}$) of \bd\ decay modes}
\label{tab:BR_Bd}
\end{table*}

\begin{table*}
{ 
\begin{tabular}{|l|rr|}
\hline
\hline
Mode	& CDF						& Belle\\
		& 1fb$^{-1}$					& 23.6 fb$^{-1}$		\\
\hline
\BsKpi	& 5.0 $\pm$ 0.7 $\pm$ 0.8 							& ($< 26$ @~90\%~CL)  \\
\Bspipi       	& 0.49 $\pm$ 0.28 $\pm$ 0.36  ($< 1.2$ @~90\%~CL)       		& ($< 12$ @~90\%~CL)  \\
\BsKK	& 											& 38 $\pm^{10}_{-9}$ $\pm$ 5 $\pm$5(f$_{s}$)  \\
\hline
\hline
\end{tabular}
}
\caption{\br\ (10$^{-6}$) of \bs\ decay modes}
\label{tab:BR_Bs}
\end{table*}

\begin{table*}
\begin{tabular}{|l|rrr|}
\hline
\hline
Mode 	& CDF 			& BaBar 						& Belle \\
		&1fb$^{-1}$		&467 M ($B\bar{B}$ pairs)		& 535 M ($B\bar{B}$ pairs) \\
\hline	 
\BdKpi	& -0.086 $\pm$ 0.023 $\pm$ 0.009         &    -0.107 $\pm$ 0.016 $^{+0.006}_{-0.004}$	& -0.094 $\pm$ 0.018 $\pm$ 0.008 	  \\
\BsKpi	&  +0.39 $\pm$ 0.15 $\pm$ 0.08			&	&    \\
\LbpK 	&  +0.37 $\pm$ 0.17 $\pm$ 0.03                     	&	&     \\
\Lbppi 	&  +0.03 $\pm$ 0.17 $\pm$ 0.05                         	&	& \\
\hline
\hline
\end{tabular}
\caption{\acp\ of \fullbhh\ decay modes}
\label{tab:ACP_Bhh}
\end{table*}

LHCb recently reported its first preliminary results for \acpbdkpi\ and \acpbskpi, using 37 pb$^{-1}$ of data \cite{acp_LHCb_vagnoni} (see \tab{ACP_LHCb}). LHC being a hadron collider, LHCb shares with CDF II some environment related characteristics, such as: high $b$-hadron production, specific trigger requirements for B-physics and, therefore, the analysis strategy is similar. In addition, LHCb benefits from the different detector structure and skills. In particular, the presence at LHCb of the RICH detectors makes possible a powerful particle identification. The efficiency in identifying the final states particles gives the possibility to measure CP-asymmetries with very competitive statistical uncertainties even if using only the small amount of data 
corresponding to about 37 pb$^{-1}$ of integrated luminosity.

\begin{table}[h]
{
\begin{tabular}{|l|r|}
\hline
\hline
Mode 	& LHCb							\\
\hline
\bdkpi\  		& -0.074 $\pm$ 0.033 $\pm$0.008 		\\
\bskpi\ 		& 0.15 $\pm$ 0.19 $\pm$0.02 			\\	   	 
\hline
\hline
\end{tabular}
}
\caption{Preliminary LHCb results for \acp\ of \bskpi\ and \bdkpi\ decay modes.}
\label{tab:ACP_LHCb}
\end{table}

\clearpage
\section{Measurement of \BsKK\ lifetime}
Using a sample of about 360 pb$^{-1}$ CDF measured the time-evolution of untagged \bskk\ decays \cite{Tonelli:2006np}. An unbiased optimization procedure, similar to the one used for \fullbhh\ analysis, aimed at improving the resolution on lifetime-measurements, yielded an offline selection based on transverse-momentum, impact-parameter, and vertex-quality requirements. The resulting signal was similar, with less statistics, to the one shown  in the left plot of ~\fig{projections}, and contained about 2200 \bskk\ decays. 
The time evolution of individual signal modes was determined by adding the decay-lenght information to the fit of composition, similar to the one described in the previous sections for the CDF analysis with 6 fb$^{-1}$.
The selection of the sample of B mesons decaying into two hadrons makes minimum requirements on the flight distance of the B meson, both during data-taking and in the final event selection. Consequently, the selection procedure tends to reject candidates which decay after a short proper time. This bias of the distribution was modeled with an efficiency curve, defined as the ratio between the pseudo-proper decay-length distribution of events passing the trigger and the unsculpted one, and it was extracted from simulations. 
The dominant sources of systematics come from effects of misalignments in the tracker, uncertainties on the model used from proper-decay time resolution and for the lifetime of background, the uncertainty of \dedx\ model, the uncertainty on the input $p_{T}(\rm B)$ spectrum used in the simulation and the uncertainty on the extraction of the trigger-efficiency curve. The resulting lifetime is $$\tau_{L}(\bskk) = 1.53 \pm 0.18 \pm 0.02~{\rm ps}.$$

\begin{figure}[h!]
\centering
\includegraphics[width=0.4\textwidth]{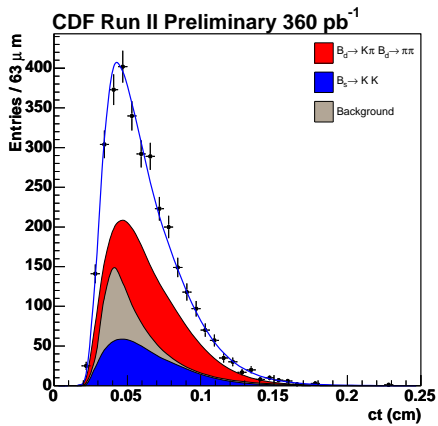}
\caption{Likelihood function projection on the proper decay length projection. The different contributions of \bd\ and \bs\ signals and of the background are overlaid.}
\label{fig:CDF_bskklifetime}
\end{figure}
The lifetime distribution of the signal with fit projection overlaid is shown in \fig{CDF_bskklifetime}.

Two independent data-driven approaches are used at LHCb to compensate for the bias of the distribution \cite{LHCb-CONF-2011-018}.
In the first one the absolute lifetime is measured by correcting for the biases by determining the event-by-event acceptance function. 
The lifetime is determined by an unbinned maximum likelihood fit using an analytical 
p.d.f. for the signal lifetime and a non-parametric estimated 
p.d.f. for the combinatorial background. The measurement is factorised in two independent fits. A first fit is performed to the observed mass spectrum and used to determine the signal and background probabilities of each event. These probabilities are used in the subsequent fit for the lifetime. The p.d.f. describing the lifetime of the signal is calculated 
analytically taking into account per-event acceptance and the proper time resolution. The p.d.f. of the combinatorial background is estimated from the data using a non-parametric method, described as a sum of one Gaussian kernel per event. The mean of the Gaussian Kernel is the measured proper time, the area of the Gaussian is weighted by the background probability of the event.
The per-event acceptance is calculated by re-running the event selection, determining  whether an event would have been selected as a function of its proper time. For example,  for an event with given kinematics, i.e. fixed track slopes and momenta, there is a direct relation between the proper time and the impact parameters of the tracks. Hence, cuts on 
impact parameters directly translate into a discrete decision about acceptance or rejection of an event as a function of its proper time.

The second approach cancels the selection bias by taking a ratio of the \BsKK\ and \BdKpi\ proper decay time distributions, exploiting the fact that \BsKK\ and \BdKpi\ have similar kinematics. 
The reliability of the procedure has been tested using a toy simulation and realistic events including the simulation of the full detector response. 
The \bskk\ lifetime is fitted for directly by means of a simultaneous fit of both final states and $\xi$ bins, where $\xi = \frac{t}{m}$ is the reduced proper time. $m$ is the invariant mass with the assignment of two pions in the final state, and therefore $\xi$ is introduced to avoid any potential bias due to a miscalculation of the candidate's proper time in the case where the final state is different from $\pi\pi$.

The dominant sources of systematics for the two measurements are contamination from other \fullbhh\ decay modes, the uncertainty on the distributions of combinatorial background, the correction of the bias, the effects of misalignments in the tracker. A systematics taking into account that in events with multiple primary vertices the B meson candidate may be assigned to the wrong vertex and hence its lifetime is wrongly calculated is also assessed.
The two measurements of the \bskk\ lifetime agree well, and using the world average measurement of the \BdKpi\ lifetime as input you can extract the \BsKK\ lifetime:
$$
\tau_{B_{s}^{0}} = 1.440 \pm 0.096  \pm 0.010 \rm ps. 
$$

\section{First search for \CP\ violation in \phiphi\ decay modes}
The $\phiphi$  decay belongs to the class of transitions of pseudoscalar mesons 
into two vector particles ($P\to VV$), whose rich dynamics involves 
three different amplitudes corresponding to the
polarization states. 
In the SM the dominant quark level process is described by the $b \to s$ penguin diagram. Hence this decay mode could provide sensitivity to the possible presence of NP in internal loops. Indeed, the SM expectation for polarization amplitudes have shown discrepancies with measurements of similar penguin decays \cite{BVV_exp}. Moreover, 
having a self-conjugate final state, the \phiphi\ allows for measuring the \bs\ decay width difference $\Delta \Gamma_{s}$, and it is sensitive to the \CP\ violation in the interference between decay with and without mixing, supplementing analogous studies in the tree dominated \bsjpsiphi\
 Actually, the \CP\ violating weak phase $\phi_s^{\phi \phi}$ is predicted to be
 extremely small in the SM  
and measurement of nonzero \CP--violating observables  
 would indicate unambiguously NP.

The first evidence for the $\bsphiphi$ decay has been reported by CDF 
in 2005~\cite{phiphi_PRL}.  Using 2.9 fb$^{-1}$ of data, the branching ratio
measurement was recently updated~\cite{phiphi},
$BR(\bsphiphi)
=(2.40 \pm 0.21 \pm 0.86 )
\times 10^{-5}$,
in agreement with the first determination.
Signal candidates are reconstructed by detecting \phikk\ 
decays and are formed by fitting four tracks to a common vertex. 
Combinatorial background is reduced by exploiting several
variables sensitive to the long lifetime and relatively
hard \pt\ spectrum of $B$ mesons, while the physics background, given by \phikst\ decay, is estimated by simulation not to exceed a
3\% fraction of the signal. Signals of $295\pm20$ events are obtained by fitting the mass distribution.
This data sample has allowed the world's first 
polarization measurement~\cite{phiphi} by analyzing the  angular distributions of decay products, expressed as a function of helicity angles,
$\vec{\omega}=(\cos\vartheta_1,\cos\vartheta_2,\Phi$). 
The total decay width is composed of three polarization amplitudes:
two \CP--even ($A_0$ and $A_\parallel$) and one \CP--odd ($A_\perp$). The measured amplitudes result in a
smaller longitudinal fraction
with respect to the na\"ive expectation, $f_{\rm{L}}=0.348\pm0.041\pm0.021$, 
 as found in other similar $b \to s$ penguin
decays~\cite{BVV_exp}.

\begin{figure} 
\begin{center}
\begin{overpic}[width=0.35 \columnwidth]{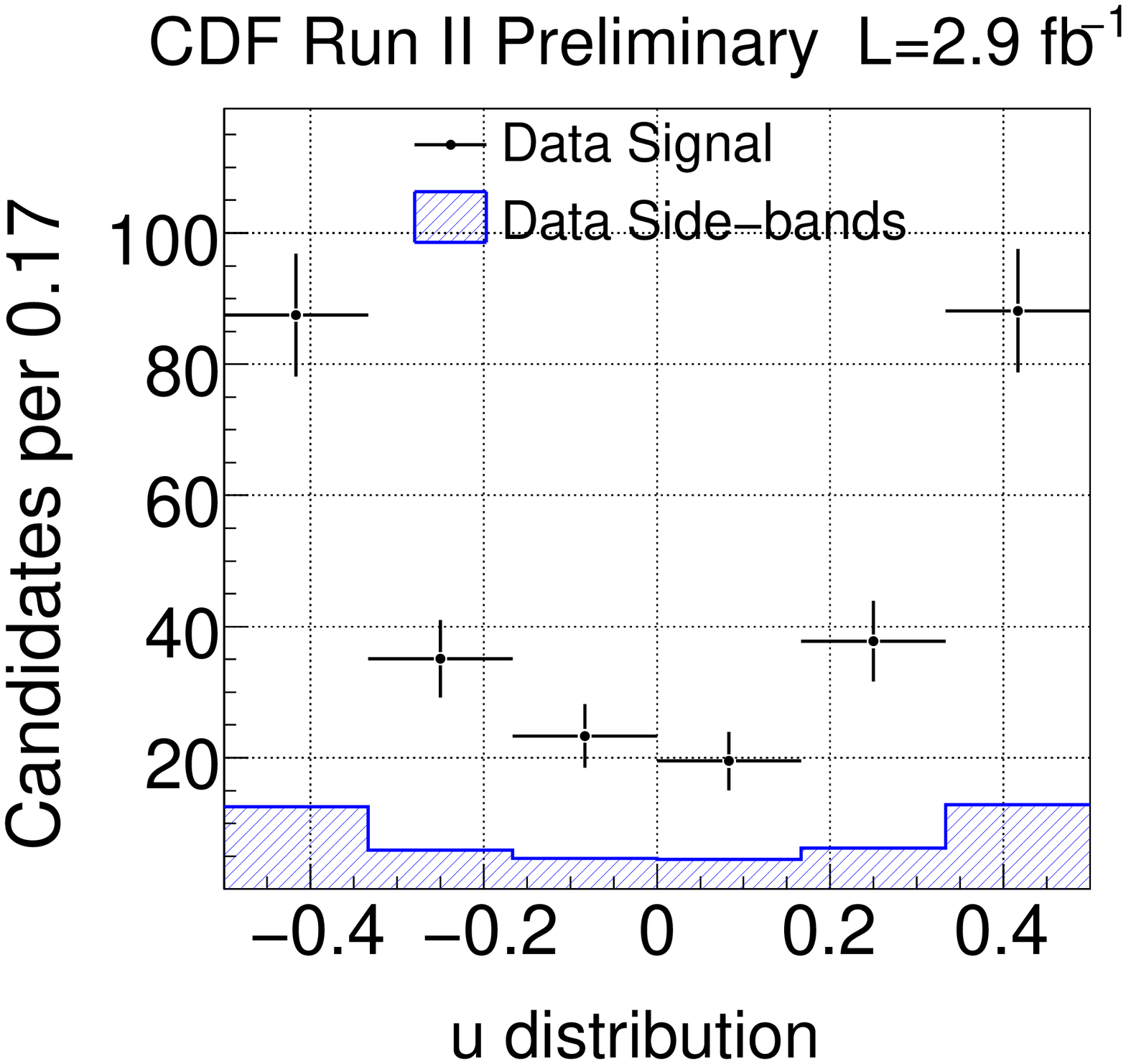}
\end{overpic}
\begin{overpic}[width=0.35\columnwidth]{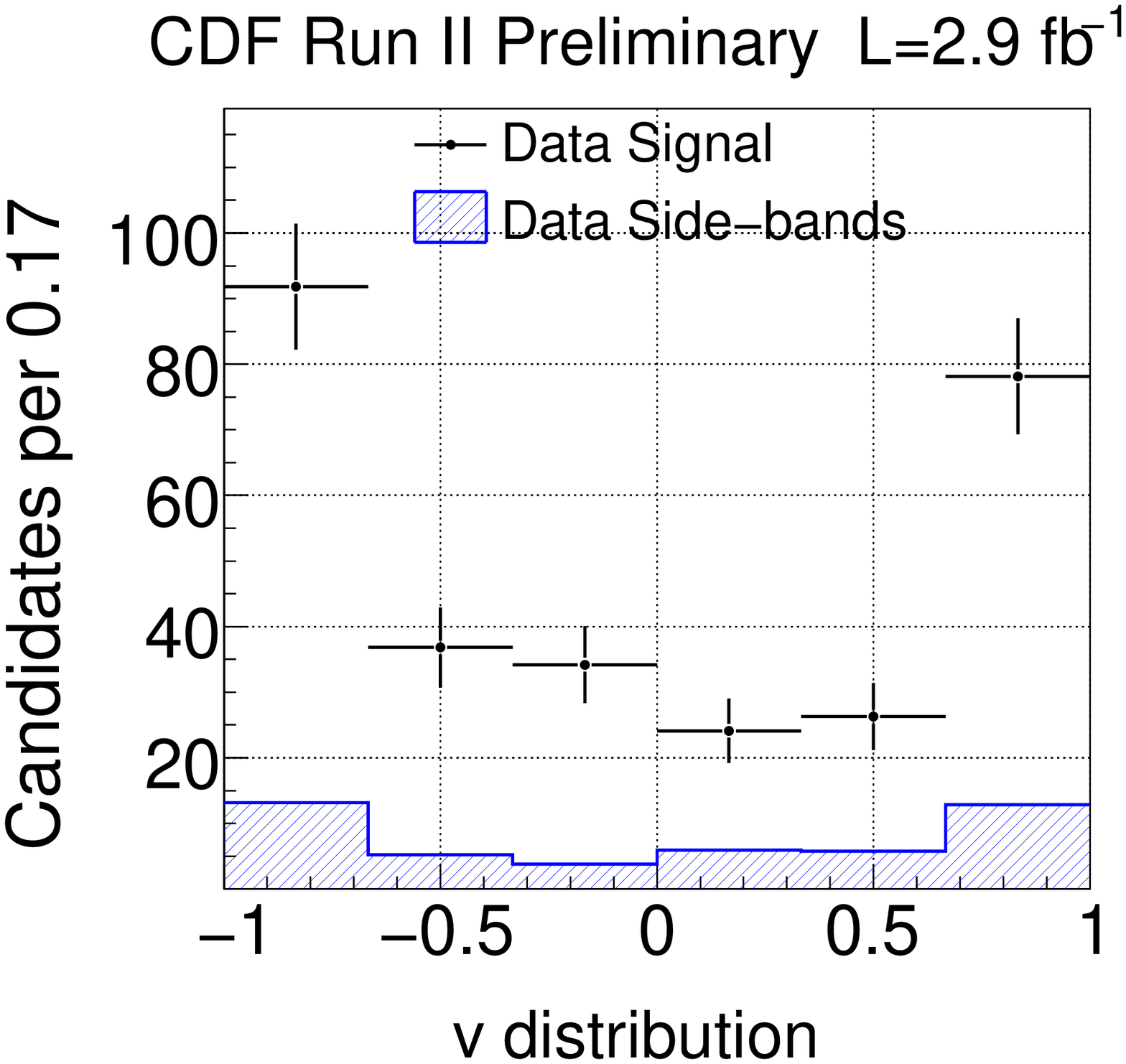}
\end{overpic}
\end{center}
\caption{\label{fig:TP}
Distribution of $u$ (left) and $v$ (right) for \phiphi\ candidates. Black crosses
are background--subtracted data; the blue histogram represents the background.}
\end{figure}

Present statistics of the \phiphi\ data sample are not sufficient for 
a suitable time--dependent analysis of mixing-induced \CP--violation as
the case of the \jpsiphi\ decay. However, an investigation of
genuine \CP--violating observables which could reveal the presence of NP, such as triple
products (TP) correlations, is accessible~\cite{TP_th}.  The TP is
expressed as $\vec{p} \cdot (\vec{\epsilon}_1 \times \vec{\epsilon}_2)$, where
$\vec{p}$ is the momentum of one of the $\phi$ meson in the \bs\ rest frame,
and $\vec{\epsilon}_i$ are the polarization vectors of the vector mesons. 
There are two triple products in the \phiphi\ decay corresponding to 
interferences between \CP--odd and \CP--even amplitudes, one for
transverse--longitudinal mixture, $\Im(A_0A_\perp^\star)$, and the other for the
transverse--transverse term, $\Im(A_\parallel A_\perp^\star)$. These
 products are functions of the helicity angles: the former is defined by
$v=\sin\Phi$ for $\cos\vartheta_1 \cos\vartheta_2\geq 0$ and $v=-\sin\Phi$
for $\cos\vartheta_1 \cos\vartheta_2<0$; the latter is defined by
$u=\sin2\Phi$. The $u$ and $v$ distribution for \phiphi\ candidates are shown in fig~\ref{fig:TP}.
 Without distinction of the flavor of the \bs\ meson at the production
time (\emph{untagged} sample), the following equation defines a \CP--violating asymmetry:
\begin{equation}
\au=\frac{\Gamma(u>0) + \bar{\Gamma}(u>0) -\Gamma(u<0) - \bar{\Gamma}(u<0)}
{\Gamma(u>0) + \bar{\Gamma}(u>0) +\Gamma(u<0) + \bar{\Gamma}(u<0)},
\end{equation}
where $\Gamma$ is the decay rate for the given process and
$\bar{\Gamma}$ is its \CP--conjugate. An equivalent definition holds
for $v$. Being proportional to $\sin\phi_s\cos\delta_i$, where $\delta_i$ are
relative strong phases between the polarization amplitudes, in
\phiphi\ these asymmetries are nonzero only in presence of NP~\cite{TP_th}.

The CDF collaboration has made the first measurement of \au\ and \av\
asymmetries in \phiphi\ using the data sample described above~\cite{TP_CDF}.
The asymmetries are obtained
through an unbinned maximum likelihood fit. 
The sample is split into two subsets
according to the sign of $u$ (or $v$) of \phiphi\ candidates. 
The invariant mass distribution of each subset is fitted
simultaneously in order to extract the signal asymmetry.  
The small fraction of physics background, such as \phikst\ as well as non--resonant decay
$\bs \to \phi K^+K^-$ and ``S--wave'' contamination \jpsifnot, is
neglected in the fit and its effect is accounted
for in the assigned systematic uncertainties. 
Using a large sample of Monte Carlo (MC) data 
the detector acceptance and the reconstruction requirements are
checked against biases
with a 0.2\% accuracy. 
The background asymmetries are consistent with zero, and the final
results for signal asymmetries are: $\au=(-0.7\pm6.4\stat\pm1.8\syst)\%$ and
$\av=(-12.0\pm6.4\stat\pm 1.6\syst)\%$. 
This measurement establishes a 
method to search for NP through \CP--violating observables in $P\to
VV$ decays without the
need of tagging and time--dependent analysis, which requires high
statistics samples. 

\section{First observation of the decay \bskstarkstar }
\begin{figure}[h!]
\centering
\includegraphics[width=0.6\textwidth]{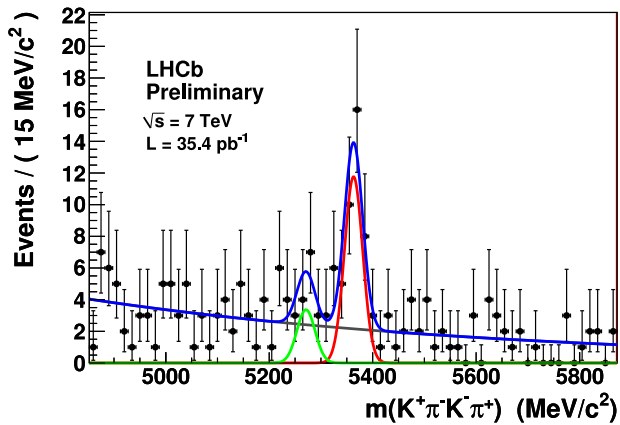}
\caption{
Fit to the $K^{+}\pi^{-}K^{-}\pi^{+}$ mass distribution for selected candidates. A gaussian function (red and green curves) is used for the \bs\ and \bd\ signals and an exponential function (blue curve) is used to describe the combinatorial background.}
\label{fig:kstarkstarmass}
\end{figure}

\bskzerostarkzerostar\ is a decay into two light vector mesons that proceeds solely through loop penguin $b\to s$ diagrams within the SM. 
The interest in \bskzerostarkzerostar\ for precision \CP\ violation studies in relation to the extraction of $\beta_{s}$ and $\gamma$ has been analyzed in \cite{bskstarstar0}, \cite{bskstarstar0a}, \cite{bskstarstar0b}.  Before the measurement we are going to present, the best measurement was an upper limit for the \BR\ of 1.68 $\times 10^{-3}$, given by the SLD experiment \cite{bskstarstar1}.
LHCb analyzed a data sample corresponding to about 35 pb$^{-1}$ of integrated luminosity. The cuts selection is optimized 
to keep maximal efficiency for the simulated decays and minimum background from a sample where no signal is expected.
The selection relies on the excellent vertexing capability of LHCb.
The mass spectrum of the selected $K^{+}\pi^{-}K^{-}\pi^{+}$ is shown in \fig{kstarkstarmass} where a clear peak is observed around the \bs\ mass.

The significance of the \bs\ signal was evaluated using the two values of the log of the maximum likelihood obtained including a Gaussian \bs\ signal and fixing the \bs\ signal to zero (null hypothesis). 
The difference between the logs corresponds to 7.4 $\sigma$ significance.
Both models include an exponential parameterization for the background, and a Gaussian signal for the \bd\ meson. When doing this test, the mass and width of the 
\bs\ meson were fixed to those obtained from an independent Þt to LHCb data from the \bsjpsiphi\ channel. Likewise, the \bd\ mass and width were input values from $B^{0}\to J/\psi \bar{K}^{*0}$
An unbinned maximum likelihood Þt was then performed, where the mass and width 
for the $B^{0}_{s}$ gaussian signal were allowed to ßoat, keeping the same exponential model to 
describe a combinatorial background. An additional gaussian signal 
was included in the model to account for a possible contribution of a \bd\ meson decay to 
the same Þnal state. The fit confirm a signal of about 34 $\pm$ 7 events for the decay \bskzerostarkzerostar.

This result can be used to provide a determination of the \br\ of \bskzerostarkzerostar\ using the normalization channel 
$B^{0}\to J/\psi \bar{K}^{*0}$ \cite{pdg_2010}, and the selection and trigger efficiencies. The other ingredients are the b$-$quark hadronization factors $f_{s}$ and $f_{d}$ \cite{Asner} to take into account the different yield of \bd\ and \bs\ mesons, and an acceptance correction.
The result is
$$
\BR (\bskzerostarkzerostar) = (1.95 \pm 0.47 \pm 0.51 \pm 0.29 (f_{d}/f_{s})) \times 10^{-5}
$$
where the systematic error is composed by 22\% uncertainty on the acceptance correction, 13\% from trigger 
efficiencies, 6\% from background subtraction and a separate error associated to the ratio $f_{d}/f_{s}$.
More details can be found at \cite{LHCb_CONF_2011_019}.


\begin{thebibliography}{99}
\bibitem{Isolation} Isolation is defined as
 $I_{B}=p_T(B)/(p_T(B)+\sum_ip_{Ti})$, where $p_T(B)$ is the transverse
 momentum of the $B$ candidate, and the sum runs over all other tracks within
 a cone of radius 1, in $\eta$-$\phi$ space around the $B$ flight-direction.
\end{thebibliography}
\end{document}